\documentstyle[preprint,aps,prb]{revtex}

\begin{document}
\draft
\title{Analysis of stability of macromolecular clusters
in dilute heteropolymer solutions}
\author{E.G.~Timoshenko\thanks{Corresponding author. 
Internet: http://darkstar.ucd.ie; 
E-mail: Edward.Timoshenko@ucd.ie}, 
Yu.A.~Kuznetsov}
\address{
Theory and Computation Group,
Department of Chemistry,\\ University College Dublin,
Belfield, Dublin 4, Ireland}
\date{\today}
\maketitle

\begin{abstract}
We study the formation of clusters consisting of several chains
in dilute solutions of amphiphilic heteropolymers.
By means of the Gaussian variational theory we show that in a region
of the phase diagram within the conventional two--phase coexistence
region {\it mesoglobules} of equal size possess the lowest free energy.
Monte Carlo simulation confirms that the mesoglobules are stabilised
due to micro--phase separation, which introduces a preferred length
scale. The very existence of such mesoscopic structures is related
to a delicate balance of the energetic and entropic terms under
the connectivity constraints. The issue of size monodispersity
and fluctuations for mesoglobules is investigated.
\end{abstract}


\section{Introduction}\label{sec:intro}

Conformational transitions in polymer solutions have been the subject
of extensive studies for many years \cite{Polymer-books}. 
In general, these are rather
complex systems with competing interactions at different ranges with
entropic contributions being equally significant. The main
progress has been made in investigating the equilibrium issues of the 
fundamental and simplest case of homopolymer solution. 
The classical Flory--Huggins theory \cite{Flory-Huggins} was further improved 
based on the scaling theory \cite{Daoud}, the self--consistent treatment 
in terms of the density variables \cite{Polymer-books} and the Lifshitz theory 
\cite{Grosb-92}.
There is also considerable experimental data available on the
phase diagrams of model systems such as polystyrene in cyclohexane 
or benzene and poly-N-isopropylacrylamide (PNIPAM) in water 
\cite{Experiment}.

However, the limit of very dilute solution appears more difficult
for experimental study. There phenomena of polymer collapse and aggregation
can go hand by hand, leading to a diverse range of theoretical
interpretations, particularly as it may be hard to separate purely
equilibrium issues from the kinetic ones  
(see e.g. discussion in Refs. \onlinecite{Chu,ItalianClusters}).
In recent attempts to resolve the controversy a considerable 
theoretical effort has been directed to understanding the collapse kinetics
of a single homopolymer \cite{deGennesKinet,GscKinet,Ganazzoli-95}. 
The necklace mechanism and some attendant kinetic laws of the collapse
transition have been obtained from the Gaussian self--consistent (GSC) method
\cite{GscHomKin}, supported in part by Monte Carlo simulations 
\cite{Rabin,CoplmMonte},
and recently reproduced using a different analytical approach in Ref. 
\onlinecite{Pitard}.

Although block and random heteropolymers have been
traditionally attracting a great deal of interest as they can exhibit
ordered micro--phase separated and disordered glassy phases
\cite{add6,add7},
their understanding is essentially limited to melts, and solutions at
high concentrations (see e.g. Refs. \onlinecite{Miles-book,Protein-book}
and references therein). It appears that the latter limitation 
is too restrictive for explaining some recent experimental findings.
In experiments on PNIPAM copolymers with small number of ionomers
in aqueous solution \cite{Deng,Qiu} it has been observed
that these polymers can form stable nanoparticles
instead of simply aggregating on heating
above the lower critical solution temperature (LCST). 
Such an unusual type of mesoscopic aggregates with extremely
monodispersed size distribution,
which we called {\it mesoglobules}, has been also reported in Ref.
\onlinecite{Gorelov} for the homopolymer PNIPAM, in which case these
structures are rather long--lived, if not truly stable. 

Clearly, the appearance of such metastable structures in homopolymer 
solutions cannot be envisaged in the framework of conventional 
Flory--Huggins type theories.
Thus, in Ref. \onlinecite{MultiChain} we have extended the GSC method
to multiple chains and argued the possibility of mesoglobules in 
dilute solution from thermodynamic considerations. 
Although the standard Flory--Huggins theory can be indeed derived
from the GSC method in the thermodynamic limit in some approximation, 
that approximation is not reliable at low concentrations.
In Ref. \onlinecite{Nuovo-Tim} we
have also mentioned the possibility that the micro--phase separation
can additionally stabilise the mesoglobules in heteropolymers, 
which as we have consequently learned from work Ref. \onlinecite{Qiu}
by Qiu et al., can take place.
 
In this paper, based on the success of the extended method of Ref. 
\onlinecite{Conf-Tra} for studying the equilibrium and kinetics of
single heteropolymers chains, we would like to support
that our conjecture by a direct numerical evidence.
Thus, it is now possible to consider several sufficiently short 
heteropolymer chains of a given composition in a box of finite volume $V$
and to analyse the values of the Helmholtz free energy ${\cal A}$ on
all its possible local minima.
We note that stationary points of the GSC equations are exactly the
extrema conditions for the variational free energy ${\cal A}$
obtained from the Gibbs--Bogoliubov principle with a quadratic
trial Hamiltonian.

Traditional approach to describing phase coexistence within
mean--field treatment relies on the following procedure.
First, one has to obtain the dependence
of the specific free energy $a=\lim {\cal A}/K$
(or possibly other equivalent thermodynamic potential) 
on the concentration $c=K/V$ (or chemical potential)
in the thermodynamic limit when $V \to \infty$ and the number of
particles $K$ diverges such that $c$ remains finite.
Second, one has to construct the convex hull of the
function $a[c]$ by applying the Maxwell construction. Within the two
phase separation region, where a straight line joining the two
free energy minima is drawn, the pure
equilibrium states are not stable with respect to fluctuations as indicated
by the wrong sign of $\partial^2 a/\partial c^2$. The 
theoretical argument that establishes the stability of the
mixed two--phase state in this case relies heavily on thermodynamic
additivity properties and that one can neglect the surface contribution of
the interface between the two phases.

Polymers, however, are pretty much finite, though normally long, chains.
Due to the connectivity of each chain there is no well defined
interface between the high and low density phases, and, moreover, 
the surface entropic contribution does seem significant too. These 
factors should be carefully accounted for while trying to study
possible metastable states in polymer solutions. 
For this end, we shall consider the system of a finite number of
finite length polymer chains, for which one cannot immediately apply 
the Maxwell construction. 
However, in this context the phase coexistence still 
means that the single--phase pure states are thermodynamically unstable.
We can expect here that there are additional metastable 
states present at the same
values of thermodynamic parameters and that strong fluctuations can 
bring the system from one of those states to another. 
Thus, the equilibrium for a finite system should be
a mixed state including a large number of local free energy minima, 
instead of just two states corresponding to the high and low density 
as in the thermodynamic limit. On increasing the system size 
we arrive at a view of the polymer precipitate as consisting of many 
various sized aggregates coexisting with a few single globules.
This picture certainly looks more adequate than 
the oversimplified mean--field inspired 
view of the phase coexistence between the two states of
one large macroglobule and a gas of single globules.

These types of problems have been extensively studied for ternary mixtures
of two immiscible liquids (water and oil) and a surfactant (e.g. diblock
copolymer) (see e.g. Refs.
\onlinecite{deGennes-Taupin,Cantor81,Micellization}),  where
metastable micelles (water in oil or oil in water surrounded by
surfactant chains)  are observed among many other more sophisticated
favourable geometrical arrangements.

In the present work we are interested in conformational structures
formed by amphiphilic (hydrophobic--hydrophilic)
copolymers in fairly dilute solutions without
presence of any third component. In this case, hydrophobic units would
tend to escape the unfavourable contacts with the solvent, but the
connectivity of each chain seriously restricts their freedom. Thus,
only micro--phase separation of both types of units within the
polymer globule is possible. For diblock copolymers we may expect
micellar globules formed with a hydrophilic shell and hydrophobic core.
At higher concentrations distinct chains may associate with each other
resulting in larger micelles consisting of several chains and the repulsive
shells of these can stop further aggregation. What type of
structures is possible for more complicated heteropolymer sequences
is not really clear as the connectivity constraints would not allow
the formation of a purely hydrophilic shell.

For heteropolymers, which possess essential  heterogeneity
along the chain, the situation seems even more complicated. 
Due to the competing hydrophobic and hydrophilic interactions 
the free energy profile is very rugged. Thus, here some new global minima
may appear  due to a specific compensation of the interactions and 
the entropy. 

Our studies based on the Gaussian variational method 
\cite{add3}
here are also supported
by direct Monte Carlo simulation, which allows us to visually
observe the system conformations and to obtain additional
insights into the problem. As kinetics after a quench to the
phase separation region is very difficult to describe reliably by
the Monte Carlo method \cite{Alan-MC}, this work deals exclusively
with the equilibrium and metastable states.

\section{Theoretical model and the Gaussian variational method}
\label{sec:model}

In this section we describe the model for any number of arbitrary
heteropolymers in solution, and introduce the Gaussian variational
method in the form derived by us in Ref. \onlinecite{Conf-Tra}.
There we have noted that the resulting equations are in fact covariant,
i.e.  their form does not depend on the structure of the
connectivity matrix between monomers.
In case of multiple chains, however, we should also include some kind of
a box, which keeps polymers from diffusing away to infinity.
This is done as in Ref. \onlinecite{MultiChain} by introducing a ``soft''
cut-off via weak ``springs'' connecting the centre--of--mass of each
chain with the centre--of--mass of the whole system.

Let us denote by ${\bf X}^a_n$ the coordinates of the $n$-th monomer
in the $a$-th chain, and to introduce multi--indices $A=(a,n)$ and so on.
Henceforth $N$ and $M$ will be the number
of monomers in a chain and the total number of chains respectively.
The effective Hamiltonian, $H$, after exclusion of
the solvent degrees of freedom is given by,
\begin{equation}\label{H}
H = \frac{k_B T}{2L^2}\sum_{a}\left( {\bf Y}^a-{\bf Y}\right)^2
+ \frac{k_B T}{2l^2}\sum_{a,n}\left( {\bf X}_n^a-{\bf X}_{n-1}^a\right)^2
+ \sum_{J\geq 2} \sum_{\{A\}} u^{(J)}_{\{A\}} \prod_{i=1}^{J-1}
\delta({\bf X}_{A_{i+1}}-{\bf X}_{A_1}),
\end{equation} 
where $L$ is the box size as in Ref. \onlinecite{MultiChain},
${\bf Y}^a \equiv (1/N)\sum_n {\bf X}^a_n$ and 
${\bf Y} \equiv (1/M)\sum_a {\bf Y}^a$ are the coordinates
of the centre--of--mass of a chain and the total system respectively,
$l$ is the statistical segment length, and $u^{(J)}_{\{A\}}$ is the set
of site--dependent virial coefficients \cite{Conf-Tra}.

The main idea of the Gaussian variational method is to use a
generic quadratic form as the trial Hamiltonian,
\begin{equation} 
H_0 = \frac{1}{2} \sum_{A,A'} V_{A\,A'}\, {\bf X}_A\,{\bf X}_{A'}. 
\end{equation}
It is possible to exclude the effective potentials
$V_{AA'}$ from the consideration and to obtain closed 
variational equations for the averages $\langle {\bf X}_A
\,{\bf X}_{A'} \rangle_0$, or, equivalently,
for the mean--squared distances between all pairs of monomers,
whether connected or not,
\begin{equation}
D_{A\,A'} \equiv \frac{1}{3}\left\langle \left( {\bf X}_A
- {\bf X}_{A'} \right)^2 \right\rangle.
\end{equation}

The trial free energy, ${\cal A} = {\cal E}-T{\cal S}$, then is obtained
according to the Gibbs--Bogoliubov variational principle,
${\cal A} = {\cal A}_0 + \langle H - H_0 \rangle_0$.
The ``entropic'' part ${\cal A}_0$ is given by \cite{Conf-Tra},
\begin{eqnarray} \label{S}
{\cal S} &=& \frac{3}{2}k_B \ln {\rm det}' R, \qquad
R_{AA'} = \frac{1}{N^2\,M^2}\sum_{BB'}D_{AB,A'B'}, \\
D_{AA',BB'} &\equiv& -\frac{1}{2}(D_{AB}+D_{A'B'}
-D_{AB'}-D_{A'B}), \nonumber
\end{eqnarray}
where the prime means that the zero eigenvalue of the matrix
is excluded from the determinant.
The mean of trial Hamiltonian $\langle H_0 \rangle_0$ is a trivial
constant and the mean energy term is given by \cite{Conf-Tra},
\begin{eqnarray} 
{\cal E} &=& 
\frac{3k_B T}{2 L^2} M\left( {\cal R}^2 
- \sum_a \frac{{\cal R}_a^2}{M} \right)
+\frac{3k_B T}{2 l^2}\sum_{n,a} D^a_{n\,n-1}
+\sum_{J=2,3}\sum_{\{A\}}\hat{u}^{(J)}_{\{A\}}({\rm det}
\Delta^{(J-1)})^{-3/2}+
 3 \hat{u}^{(3)}\sum_{A\not=A'} D_{AA'}^{-3}, \label{E} \\
&&\Delta^{(J-1)}_{ij} \equiv D_{A_1 A_{i+1},A_1 A_{j+1}}, \qquad
\hat{u}^{(J)}_{\{A\}} \equiv (2\pi)^{-3(J-1)/2} u^{(J)}_{\{A\}},
\nonumber
\end{eqnarray}
where we have included the volume interactions up to the 
three--body terms only, so that $u^{(J)}_{\{A\}}=0$ for $J>3$,
and for the discussion of the last term see Ref. \onlinecite{Qz-mess}.
In Eq.~(\ref{E}) the total and partial radii of gyration are defined
as follows,
\begin{equation}
{\cal R}^2 = \frac{1}{2N^2M^2}\sum_{AA'}D_{AA'}, \qquad
{\cal R}_a^2 = \frac{1}{2N^2}\sum_{nn'}D^{aa}_{nn'}.
\end{equation}

We shall use the following particular parametrisation for the matrix
of the second virial coefficients,
\begin{equation} \label{gsc:u2hh}
u^{(2)}_{AA'}=\bar{u}^{(2)}+\Delta\frac{\sigma_A+\sigma_{A'}}{2}.
\end{equation} 
This corresponds to the case of amphiphilic heteropolymers, for which
monomers differ only in the monomer--solvent coupling constants.
Then the mean second virial coefficient, $\bar{u}^{(2)}$, is associated
with the quality of the solvent and the parameter $\Delta$ is called the
degree of amphiphilicity of the chain.   
The set $\{\sigma_n\}$ expresses the chemical composition, or the
{\it primary sequence} of a heteropolymer chain. In our case the
variables $\sigma_A$ can take only two values:
$-1,1$ corresponding to the hydrophobic `$a$' and hydrophilic `$b$'
monomers respectively.

It is worthwhile to comment on the origin of this parametrisation
linear in the composition variables.
One usually proceeds from the effective Hamiltonian,
\begin{equation}
H_{ms}=H_{solv}[{\bf R}_a] + H_{mon}[{\bf X}_A]-
\sum_{n,\alpha} I_A \,\delta({\bf X}_A -{\bf R}_{\alpha}),
\label{EfEf1}
\end{equation}
which includes the terms describing the solvent degrees of freedom, 
${\bf R}_{\alpha}$, the monomer degrees of freedom, and a `contact' 
monomer--solvent interaction, characterised by the
$A$-th monomer hydrophobic strengths, $I_A$,  respectively. 
A simple way
of deriving such a term, proposed by Garel and Orland \cite{add6},
would be then to explicitly use the 
solution incompressibility condition, 
\begin{equation}
\rho_{mon}({\bf y}) +\rho_{solv}({\bf y})=
\sum_A \delta({\bf y}-{\bf X}_A) +
\sum_{\alpha} \delta({\bf y}-{\bf R}_{\alpha})=\rho_0 = const,
\label{IncCond}
\end{equation}
in order to integrate
out the solvent degrees of freedom. This yields the partition
function $Z_{ms} = Z_{solv}\,Z$, where the effect of the solvent influence
on the monomer degrees of freedom appears in $Z$ only via
the following term in the effective Hamiltonian, 
\begin{equation}
H=\sum_{A\not= A'}\left( u_2 +\frac{1}{2}(I_A+I_{A'})\right)\,
\delta({\bf X}_A -{\bf X}_{A'})+ \ldots
\label{EfEf2}
\end{equation}
Now, by introducing $\bar{u}_2=u_2+I$ and $\sigma_A=I_A-I$, where
$I$ is the mean value of $I_A$, we obtain exactly the linear term. 

We note that the widely used quadratic term in the composition variables
corresponds to the Edwards
free energy functional constructed as the virial expansion,
$\sum_L \int d{\bf y}\,(\rho_p({\bf y}))^L$, in terms of
the pseudo--density, $\rho_p({\bf y})=\sum_{A}\sigma_A\,
\delta({\bf X}_A -{\bf y})$. This model was widely exploited 
\cite{add7}, and although it is clearly suitable
for a mean--field  theories, microscopically it corresponds
to rather non--local monomer--solvent interactions.
The model with the quadratic term in composition
variables, often called the random `charge' model, 
is appropriate for describing either true charges or non--local
effective monomer--solvent interactions arising after coarse--graining
of models with complex intra--molecular potentials.
It is popular for the use in studying protein folding
because proteins are too complicated to be described well
by any model.

To be specific, we also choose to fix the third virial coefficient
$u^{(3)}_{mm'm''}=10\ k_B T l^6$.
As usual, we work in the system of units such that $l=1$,
and $k_B T =1$.

\section{Lattice model and the Monte Carlo technique}

For simulation we adopt the Monte Carlo technique in the lattice model of 
heteropolymers from Ref. \onlinecite{CoplmMonte}.
Thus, apart from the connectivity and excluded volume constraints there
are ``weak'' pair--wise interactions between
lattice sites depending on the separation and sites contents,
described by the Hamiltonian,
\begin{equation}
   H = \frac{1}{2} \sum_{i \ne j} w (r_{ij})
       {\cal I}_{s_{i}s_{j}},
       \label{lmc:hamil}
\end{equation}
where $i$, $j$ enumerate lattice sites; $s_{i}$ labels the contents of
site $i$, ${\cal I}_{s_{i}s_{j}}$ is a $3\times 3$ symmetric
matrix of monomer and solvent interactions, 
$r_{ij} = \vert{\bf r}_{i} - {\bf r}_{j}\vert$ 
is the separation between the two sites, and
$w (r_{ij})$ is a function giving the shape
of the potential.

The lattice model, similarly to Eq.~(\ref{IncCond}), describes an
incompressible solution --- each site which is not occupied by a monomer
contains a solvent molecule.
The Hamiltonian (\ref{lmc:hamil}) can be rewritten in the equivalent form,
\begin{equation}
H = \sum_r w(r)\left(
    {\cal I}_{aa} C_{aa}(r) + {\cal I}_{bb} C_{bb}(r) +
    {\cal I}_{ss} C_{ss}(r) + {\cal I}_{ab} C_{ab}(r) +
    {\cal I}_{as} C_{as}(r) + {\cal I}_{bs} C_{bs}(r) \right),
\label{lmc:Hascont}
\end{equation}
where $C_{lm}(r)$ is the total number of $lm$-contacts at the $r$-th
interaction range.
Such a contact is defined as a pair of lattice sites at the distance $r$
occupied by `l' and `m' species.
In this model \cite{CoplmMonte} we include first nearest neighbours
($w(r = 1) = 1$), 2D and 3D diagonals ($w(r = \sqrt{2}) = 1$ and
$w(r = \sqrt{3}) = 0.7)$, as well as the second nearest neighbours
($w(r = 2) = 1/2$), so that no higher interaction ranges are present
($w(r > 2) = 0$).
Due to solution incompressibility numbers of both types of monomers
and solvent molecules are fixed. This yields additional constraints
on the number of contacts.
Indeed, by considering contacts formed by `a'-monomers, we can write,
\begin{equation}
2 C_{aa}(r) + C_{ab}(r) + C_{as}(r) = {\cal C}(r) N_a, \label{lmc:acontact}
\end{equation}
where $N_a$ is the total number of `a' monomers on the lattice
and the factor ${\cal C} (r)$ is the total number of contacts at
the $r$-th interaction range per lattice site
(in our case, ${\cal C} (1) = 6$, ${\cal C} (\sqrt{2}) = 12$,
${\cal C} (\sqrt{3}) = 8$ and ${\cal C} (2) = 6$).
Analogously, by considering `b' and `s' lattice
sites we can write respectively,
\begin{eqnarray}
2 C_{bb}(r) + C_{ab}(r) + C_{bs}(r) & = & {\cal C}(r) N_b,
  \label{lmc:bcontact}\\
2 C_{ss}(r) + C_{as}(r) + C_{bs}(r) & = & {\cal C}(r) (L^3 - N_a - N_b).
  \label{lmc:wcontact}
\end{eqnarray}
Using relations (\ref{lmc:acontact},\ref{lmc:bcontact},\ref{lmc:wcontact})
one can totally exclude the monomer--solvent contacts
from consideration similarly to Eqs.~(\ref{EfEf1}, \ref{EfEf2}) and
rewrite the Hamiltonian (\ref{lmc:Hascont}) as follows,
\begin{eqnarray}
H & = & H_0 - k_B T \sum_r w(r) \left( \chi_{aa} C_{aa}(r) +
    \chi_{ab} C_{ab}(r) + \chi_{bb} C_{bb}(r) \right), \label{lmc:chihamil} \\
H_0 & = & ({\cal C}/2) L^3 {\cal I}_{ss} +
          {\cal C} N_a ({\cal I}_{as} - {\cal I}_{ss}) +
          {\cal C} N_b ({\cal I}_{bs} - {\cal I}_{ss}),\quad
          {\cal C} = \sum_r w(r) {\cal C}(r). \label{lmc:junkhamil}
\end{eqnarray}
Here we have introduced the so--called Flory interaction parameters,
\begin{equation} \label{lmc:chiab}
\chi_{aa} = \frac{2 {\cal I}_{sa} - {\cal I}_{aa} -
                {\cal I}_{ss}}{k_{B}T}, \quad
\chi_{bb} = \frac{2 {\cal I}_{sb} - {\cal I}_{bb} -
                {\cal I}_{ss}}{k_{B}T}, \quad
\chi_{ab} = \frac{  {\cal I}_{sa} + {\cal I}_{sb} -
                {\cal I}_{ab} - {\cal I}_{ss}}{k_{B}T}.
\end{equation}
The first term in (\ref{lmc:chihamil}) is just trivial constant
which does not depend on the system conformation and can be neglected.
The combinations of interaction parameters in
Eq.~(\ref{lmc:chiab}) describe the degree of
corresponding monomer--monomer attraction and they are the
only relevant thermodynamic parameters characterising interactions
in the system for a given number of $M$ polymer sequences of length
$N$ and lattice size $L$.

Thus, as we have seen, the Hamiltonian (\ref{lmc:chihamil}) possesses
a similar structure to the effective Hamiltonian in Eq.~(\ref{H})
from the Gaussian theory of the previous section.
The minor distinction is that in
the lattice model the connectivity and excluded volume constraints
are explicitly implemented \cite{CoplmMonte}.
The relation of each of the Flory parameters to the virial coefficients
then can be worked out similarly to the derivation of the standard
Flory--Huggins theory \cite{Polymer-books}:
$u^{(2)}_{lm} \sim l^3 ( const - 2\chi_{lm})$,
$u^{(3)} \sim l^6$, $\ldots$,
where $l$ is the lattice spacing and the $const$ depends on the particular
choice of $w(r)$ weight function only.
Given the latter relations, Eqs.~(\ref{EfEf2}) and (\ref{lmc:chihamil},
\ref{lmc:chiab})
differ only by replacing the Dirac delta--functions to contacts via the
Kronecker symbols on the lattice.
Finally, parametrisation of the second virial coefficients for amphiphilic 
heteropolymers Eq.~(\ref{gsc:u2hh}) in the present model results in an
additional relation \cite{CoplmMonte},
$\chi_{aa}+\chi_{bb}=2\chi_{ab}$.

For finding equilibrium and metastable states 
one is free to use a combination of various 
Monte Carlo moves which relax the system faster.
In addition to local monomer moves \cite{CoplmMonte}, we argue that
for a multichain system it is highly 
desirable to include translational moves of whole chains. 
This can be motivated as follows.
Once a polymer has collapsed the chain mobility deteriorates
significantly in the scheme with local moves only. 
Indeed, monomers forming a space filling core of the globule can hardly
move at all, and movements of the globule shell contribute little 
to translations of the globule. Thus, aggregation
and collapse would become oversuspended.

The situation improves dramatically by introducing
translational moves representing the diffusion of chains.
To be consistent, however,
clumps involving several polymers should be treated by the scheme of 
translational moves in exactly
the same manner as single chains.
Thus, translational moves are applied to all {\it clusters}
of chains within the interaction range 
with a probability inversely proportional to
the number of monomers within. This ensures the Stokes law in the
absence of the hydrodynamic interaction. 
Such a translational move results in shifting the current 
cluster in a random direction among 6 possible directions.

\section{Homopolymer solution}

We precede the main results by considering the homopolymer solution,
for which all minima of the free energy corresponding to clusters
are expected to be unstable according to the standard theories.

The phase diagram of the homopolymer solution can be easily
obtained from the variational equations by using the additional
kinematic assumption in case of ring polymers \cite{MultiChain}
that the mean squared distances between any two
monomers belonging to two distinct chains are the same
$D_{mm'}^{aa'}\equiv \overline{D}$, and may be written
as $D_{mm'}^{aa}\equiv D_{|m-m'|}$ for any two monomers belonging to 
the same chain. We have explicitly checked that in the more general
formalism described here these assumptions are automatically satisfied
for the thermodynamically stable states, i.e. the main minima of ${\cal A}$,
for solution of ring homopolymers. 
In Ref. \onlinecite{MultiChain} we have concluded that
the boundary of the coexistence region is well described by
the Flory--Huggins theory.
In fact, Eqs. (14,15,17) in the simplified treatment
of Ref. \onlinecite{MultiChain} are
capable to describe only the ``symmetric'' phases, for which
all polymer chains stay apart from each other or collapse
into a single precipitate. 
However, for the metastable states,
i.e. local free energy minima, this is not true due to the phenomenon of
spontaneous symmetry breaking analogous to our discussion in Ref.
\onlinecite{Conf-Tra}.

Numerical analysis of the complete
set of variational equations shows that there are
additional states related to various local free energy
minima. These minima correspond to conformations where
polymers in solution form several clusters, each consisting
of one or a few chains. Obviously, for a large number of chains there
may be many such states.
The situation is illustrated in Fig.~\ref{fig:mh4}, in which we present
the mean squared radii of gyration of clusters formed by
various number of chains in solution of  $M = 4$ homopolymers.
The lines are drawn as long as the corresponding minima of the
free energy exist. One can see from Fig.~\ref{fig:mh4} that
for rather small number of chains such ``asymmetric'' minima
exist in a certain region of the phase diagram bounded   
by the first point of curve $4\times 1$ on the left and the last point
of curve $1\times 4$ on the right \cite{Footnote0}. Moreover, the upper bounds
in $u^{(2)}$ for the existence of such states are approximately the same.

Now let us compare the values of the free energy at
minima corresponding to various possibilities
to divide the system into subsets.
In Tab.~\ref{tab:many_hom} we present values of the
free energy for some cluster sizes, corresponding to
either a symmetric division of the system into clusters of
equal size, or a large precipitate plus one or two single
globules. One can see from Tab.~\ref{tab:many_hom} that
all asymmetric minima for the homopolymer are
not stable. Note also that the system divided into
a large aggregate plus a few single globules possesses
the free energy value close enough to that at the global
minimum, whilst the system divided into several clusters
of equal size has a higher free energy.
In fact, one would expect that for the homopolymer solution
of a huge number of chains the minimum of the free energy
should be at one of the asymmetric states consisting
of one large precipitate plus a gas of single globules.
Such behaviour would be consistent with the standard
picture of two--phase coexistence in the thermodynamic
limit. However, due to high computational expenses
we have not been able to test this properly
based on the variational equations as yet.

\section{Heteropolymer solutions}

\subsection{Results from the variational method}

First, let us overview the main results obtained by numerical
analysis of the complete set of the extrema conditions of 
the variational free energy Eqs. (\ref{S},\ref{E}).
In Figs.~\ref{fig:mcp_1a1b_phd} and \ref{fig:mcp_3a3b_phd} we present
the equilibrium phase diagrams for solution of $M=4$ heteropolymers
consisting of short and long blocks respectively.
These diagrams are drawn at a fixed concentration in terms
of the mean second virial
coefficient $\bar{u}^{(2)}$ and the amphiphilicity $\Delta$,
which parametrise the matrix of the two--body virial coefficients
in Eq.~(\ref{gsc:u2hh}).
For small values of the amphiphilicity $\Delta$ the phase diagrams
of heteropolymers are essentially the same as for the homopolymer.
Thus, there are the low--density phase of individual globules
(or coils in the repulsive regime) and the high--density macroglobule, 
as well as the region of their coexistence.

Let us now discuss how the situation changes with increasing
$\Delta$ at a fixed low concentration. 
For the two heteropolymers under consideration there appears an
intermediate region, in which the state corresponding to two
clusters of two chains each possesses the lowest free energy.
Such a minimum appears starting from some critical value of the
amphiphilicity and is bound to a narrow range in the mean
second virial coefficient for any fixed $\Delta$.
As it is clear from Figs.~\ref{fig:mcp_1a1b_phd} and 
\ref{fig:mcp_3a3b_phd}, this region designated as the `Mesoglobules'
expands with increasing the amphiphilicity.
The location and shape of this region turn out to be very sensitive on
the heteropolymer sequence. For long blocks heteropolymers this region
is narrower and appears at a weaker attraction, characterised
by a smaller $|\bar{u}^{(2)}|$ compared to the case of
short blocks (alternating monomers).
Indeed, for the former the micro--phase separation, which stabilises
the mesoglobules, proceeds easier, i.e. it requires a weaker attraction
to occur. It is important to emphasise that the asymmetric
clusters `$3+1$' and `$1+1+2$' always possess a higher 
free energy than other minima (i.e. the macroglobule `$1\times 4$', 
the single globules `$4\times 1$' or the mesoglobules `$2+2$'),
and thus are merely metastable.

Now then, let us examine somewhat larger systems composed of
$M=12$ chains of length $N=12$ with varying block length.
Values of the free energy at various local minima, corresponding to
symmetric and some asymmetric clusters, are presented in the series
of Tabs. \ref{tab:mcp_1a1b}, \ref{tab:mcp_2a2b}, \ref{tab:mcp_3a3b} and
\ref{tab:mcp_rand}
for different values of $\bar{u}^{(2)}$ at a fixed sufficiently
high $\Delta$ for different periodic and aperiodic sequences. 
All considered asymmetric clusters have been
found to possess a higher value of the free energy than the
symmetric ones. The main conclusion from the above case of
a smaller system that the mesoglobules are thermodynamically
stable in some intermediate region, remains valid here as well.
However, in the current case a few different mesoglobules sizes are possible,
namely `$6\times 2$', `$4\times 3$', `$3\times 4$' and `$2\times 6$'.
We also find that at a given mean second virial coefficient, amphiphilicity,
concentration and fixed sequence, only one of these is thermodynamically stable.
With all other parameters fixed, the size of stable mesoglobules
increases with the concentration and 
with $|\bar{u}^{(2)}|$.
Thus, the equilibrium transition from the gas of single
globules to the macroaggregate on increasing $|\bar{u}^{(2)}|$
proceeds in a few steps.
Clearly, the number of various possible clusters grows exponentially
with the system size, and for a sufficiently large system it is
impossible to enumerate all possible divisions. 
We emphasise that according to Tab.~\ref{tab:mcp_rand} symmetric clusters
have the lowest free energy not only for heteropolymers with a periodic
block structure, but essentially for many aperiodic sequences as well.

\subsection{Results from lattice Monte Carlo simulation}\label{ResMC}

It is interesting to check these predictions of the Gaussian
variational method by the Monte Carlo simulation on a lattice.
Here we shall consider several concrete sequences of amphiphilic heteropolymers
consisting of strongly hydrophobic and slightly hydrophilic units such that
$\chi_{aa} = -5 \chi_{bb}$.
We fix the main parameters as follows:
linear lattice size $L=60$, polymer length $N=24$ and number
of chains $M=20$. To obtain final equilibrium states we proceeded
from a random coil state ($\chi_{aa}=0.1$) and performed an instantaneous 
quench to $\chi_{aa}=1$ followed by a few millions of Monte Carlo sweeps of 
relaxation  \cite{Footnote1}. 

Fig.~\ref{fig:nclvt} shows the time
evolution of the mean number of macromolecular clusters $n_{cl}$ 
during the relaxation.
The solid curve in this figure corresponds to the homopolymers consisting
of only hydrophobic units, for which the final equilibrium 
is reached after a few hundred thousands of MC sweeps. The resulting
state is a single aggregate of $20$ chains, as the coexisting low
density phase of single globules is virtually unobservable here due
to the nearly vertical shape of the left two--phase coexistence boundary
in the Flory--Huggins phase diagram. In the $n_{cl}(t)$
dependence for heteropolymers first there is
a similar fast stage, which is then followed by an extremely
slow further relaxation.
Essentially no change in the value of $n_{cl}$ happens at
very large times, which shows that the final
equilibrium has indeed been reached for considered sequences.
Remarkably, the number of clusters in the final state here is
not equal to unity, and attempts to carry on the simulation further
never changed the situation.

In Figs. \ref{fig:SnaPgoodNbad} and \ref{fig:SnaPmeso}
we exhibit snapshots of typical system
equilibrium conformations for different heteropolymer sequences.
In Fig. \ref{fig:SnaPgoodNbad}a we have a snapshot
of the initial state of swollen coils at very weak overall monomer attraction
insufficient to overcome the entropic effect.
In Fig.~\ref{fig:SnaPgoodNbad}b we have a snapshot of the final
single aggregate in the case of diblock copolymers, which has
a clear micellar structure of a hydrophobic core (black)
and a shell of hydrophilic (white) subchains sticking out.

Snapshots in Figs. \ref{fig:SnaPmeso}a--\ref{fig:SnaPmeso}d correspond
to the final states of the system for different heteropolymer sequences.
These correspond to a few distinct clusters each consisting of several chains,
which have a large amount of hydrophilic (white) material on the outside.
Strikingly,  in case of sequences in 
Figs. \ref{fig:SnaPmeso}b--\ref{fig:SnaPmeso}d
these clusters have nearly equal size, i.e. well monodispersed particles
are produced. We have already seen from the Gaussian variational theory
that conformational structures corresponding to clusters of equal
size (which we called symmetric there) may become most stable in some area
of the phase diagram.
Now, on the lattice,
a direct computation shows that these mesoglobules possess
a somewhat higher energy than the single macroglobule at the same 
values of interaction parameters. However, their entropy is, 
obviously, higher as well due
to the gain of translational entropy,
and the net result in the free energy favours the mesoglobules,
which is manifested in their apparent stability. This interplay of different
contributions is rather subtle, and, clearly, the micro--phase separation,
which leads to a repulsive shell on the surface of the mesoglobules, does
play a significant role. 

Now let us examine the question about the size polydispersity of
the mesoglobules in more detail. For this we have performed the
above described relaxation procedure for a large ensemble consisting
of $Q=1000$ independent different initial conditions. In 
Figs.~\ref{fig:MclHys} and \ref{fig:RclHys} we present the calculated
histograms of the mass (i.e. number of chains in a mesoglobule) 
and size (i.e. squared radius of gyration of a mesoglobule) distributions
in the final state for different sequences.

The most typical picture is seen for sequences s2, s3 (intermediate sized
blocks) and s6 (an irregular (randomly generated) sequence). 
These have a single well distinguished
peak in the mass and size distributions, which has a Gaussian--like
shape with a fairly narrow width. This corresponds to
essentially monodispersed mesoglobules, which for our
particular system size have about 10-15 percent relative dispersion in
linear size. 
Some sequences, however, do not result in monodispersed mesoglobules.
For example, sequence s1 (alternating very short blocks) has 
two peaks in its mass distribution: a large population
of single globules $M_{cl}=1$ and a smaller population of mesoglobules
consisting of about 2-6 chains. A typical snapshot for this
sequence in Fig. \ref{fig:SnaPmeso}a has two single globules and
two large formations of fairly irregular shape and different size.
The main reason for this is that due to a very short block length 
forming a core and shell structure is not possible.

The mass distribution for sequences s4 (diblock copolymer) and
s5 (inverse to the irregular sequence s6, which has mostly
hydrophobic ends) in addition to a mesoglobules
peak possess a large population of single aggregates
$M_{cl}=20$. The latter circumstance is due to that the characteristic size
of mesoglobules here (about 15 for s4) is quite large and comparable to the 
number of chains $M=20$. Thus, we may expect that even for these
sequences with increasing the system size (i.e. $M$ and $L$ so that 
$c=M/L^3=const$) these two peaks would transform to a single mesoglobules
peak as for sequences s2, s3, s6. However, to prove this reasonable
conjecture by simulation would require enormous computational times
\cite{Footnote2}.
Note that, at the same time, 
the size distribution for s4, s5 has essentially a single 
fairly narrow peak even for $M=20$.

Thus, we see that the size of mesoglobules varies within a few dozens
percents margin due to fluctuations. This situation is analogous to
that of the Gaussian variational theory, in which clusters of slightly unequal
sizes have close, but somewhat higher free energy than the respective
symmetric clusters. This means also that the barriers separating
such slightly different minima make it hard for the system to
transform from one of the metastable states to the true free energy
minimum. 
In Monte Carlo simulation fluctuations permit to move a single
chain from one cluster to another occasionally, but the average mesoglobules
size does not really fluctuate.
Transitions between states with different mean size of mesoglobules
are strictly suppressed due to rather high activation barriers.
Finally, we remark that
adequate description of the nucleation process would require
an introduction of collective moves, which can split clusters 
and form new ones, to the Monte Carlo scheme.
Thus, we do not attempt to describe any dynamic or kinetic properties
of heteropolymer solutions in this work.

\section{Conclusion}

In this paper we have studied the equilibrium conformational
states in solutions of amphiphilic heteropolymers at relatively low
concentrations.
The main conclusion from our considerations is that in heteropolymers
there are additional thermodynamic states obtained by association of several
distinct chains. 
This effect is specific to heteropolymers with sufficiently strong
competing interactions. In homopolymer solution clusters of
several chains always possess a higher free energy than the
gas of single globules or the precipitate, so that such states 
cannot be stable.
We have introduced the term {\it mesoglobules} to refer to
rather monodispersed (or exactly equal sized in the mean--field
approximation) mesoscopic globules, i.e. globules composed
of more than one and less than all chains.

The average size of the mesoglobules in heteropolymer solutions is
determined by the characteristic scale of the micro--phase separation.
The physical mechanism responsible for it has been discussed in 
Sec.~\ref{ResMC}. Formation of mesoglobules from a single macro-aggregate
at the same values of interaction parameters results in:
a) a favourable gain of translational entropy,
$\Delta {\cal S} \simeq n_{mes}\,\ln (V/n_{mes}) - \ln V$,
where $n_{mes}$ is the number of mesoglobules;
b) an unfavourable gain of surface energy 
$\Delta {\cal E} \simeq \varsigma_1(\Delta, \sigma_i)\, n_{mes}\, R^2_{mes}
- \varsigma_2(\Delta, \sigma_i)\, R^2_{mac}$, where the mean radii of 
a mesoglobule and the macroaggregate can
roughly be  estimated as 
$R_{mes} \sim \left(N M u^{(3)}
/(n_{mes} |\bar{u}^{(2)}|) \right)^{1/3}$ and 
$R_{mac} \sim \left(N M u^{(3)}
/|\bar{u}^{(2)}| \right)^{1/3}$
respectively. Here $\varsigma_{1,2}(\Delta,\sigma_i)$ are some `effective'
surface tension coefficients which arise from a rather complicated
mismatch between the amounts of hydrophobic and hydrophilic
units exposed on the surface for a given sequence in the two cases.
These two tendencies compete with each other, but it is
more favourable to produce mesoglobules of certain size
in a rather narrow domain of the phase diagram as is
seen e.g. in Figs. 2,3 and Tabs. II and III. 
In the case of periodic copolymers with fairly long blocks it is clear that
a large scale phase separation of `a' and `b' units would play a major role
for both forming mesoglobules in dilute solution and a shell--and--core
single globule at infinite dilution \cite{Orlandini,Conf-Tra}.
Perhaps, in case of more complicated irregular sequences a sort
of more refined Imry--Ma argument \cite{ImryMa}, which would take into
account the above described balance of energic and entropic terms,
may provide further insight into formation of mesoglobules as a 
kind of localised
domains appearing due to the coupling 
$\sum_{a,i} \sigma_i\, \rho_{mon}({\bf X}_i^a)$
of the monomer density 
to `disordered' variables $\sigma_i$ in the Hamiltonian Eq. (\ref{H}).

The size distribution of the mesoglobules is sufficiently monodisperse
due to a thermodynamic preference for clusters to be of equal
size. However, fluctuations can transform symmetric clusters
into a slightly asymmetric ones, although the barriers separating these
structures from strongly asymmetric clusters, such as 
macroaggregates, are very high. 

We find that short blocks and certain `good' irregular sequences also
form mesoglobules in some narrow regions of the phase diagram. 
The conformation of these mesoglobules cannot have a clear 
micellar structure due to the connectivity constraints 
which make it very difficult to form a core of hydrophilic 
and a shell of hydrophobic units for a given sequence.
Other sequences, such as some
`bad' aperiodic sequences and, as we have seen, in our regime also
diblocks, which may be viewed as model surfactants, 
can only produce particles with a broad size distribution.
Some of more complicated anomalous cases may nevertheless be quite
interesting. So, for example, sequence s5 in Fig. 7b
(but, significantly, not s6 which is obtained by `a' to `b' mutual
replacements), 
which contains essentially two hydrophobic end blocks with a hydrophilic
block in the middle, produces a quite polydispersed cluster distribution.
Snapshots of corresponding conformations show a number of clusters
interconnected by short hydrophilic bridges.
These, of course, do not qualify as mesoglobules.
Clearly,
at higher concentrations, this localised network formation would play
an increasingly important role.
We hope to be able to return to the study of conformational structures
produced by such tri--block and more peculiar sequences
in dilute solutions at a later date.

What also seems to be essential for the existence of the mesoglobules is that
there should be sufficient distinction in monomer--solvent
interactions between the two types of units,
one of which should be hydrophobic and another slightly hydrophilic.
No mesoglobules have been found by us for hydrophobic--neutral
heteropolymers (for which $\chi_{aa} > 0$ and $\chi_{bb} = 0$),
which tend to simply aggregate similar to the homopolymer case.

Our formal observation in this work was that the mesoglobules are
mean--field stable (rather than merely metastable)
states in some regions of the phase diagram. However, due to fluctuations
beyond mean--field they are not fully monodisperse, but possess a fairly narrow
size distribution. Nevertheless, no matter how much more time elapses
they do not tend to grow or aggregate and preserve their mean size
and distribution. The latter conclusion is supported by our Monte Carlo
simulations on extremely long times \cite{Footnote2}
and seems to be in agreement
with experimental evidence \cite{Gorelov}.
We thus believe that the observed phenomenon is generic for
fairly dilute heteropolymer solutions.
As for the exact regions of stability of mesoglobules and their mean
size and monodispersity, these seem to be extremely sensitive on the
particular heteropolymer sequence and thermodynamic parameters of the
system.
It also seems quite feasible that even a very weak electrostatic repulsion
may play crucial role for further stabilisation of mesoglobules
and that it can improve their monodispersity significantly.

Unfortunately, at the moment no theory exists that could describe
the dynamic and kinetic phenomena in heteropolymer solutions at
the same level of detail as we have been able to achieve
here for the equilibrium and metastable states.
The Gaussian self--consistent method is an optimised 
mean--field type theory and the
account for nucleation and density fluctuations
is beyond its scope.
On the other hand, the Monte Carlo method is difficult to apply to kinetics as
the issue of choosing a particular scheme of Monte Carlo
moves is obscure. Besides, no simulation alone can completely convincingly
distinguish between true thermodynamic
states and very long lived metastable ones.

However, it seems that even the limited information on the depths
of various local minima and barrier heights between them
obtained from the Gaussian variational method should be sufficient
for understanding 
and explaining some novel phenomena observed in recent experiments with
heteropolymer solutions. For instance, it is possible that in
kinetic experiments the size of mesoglobules would be dependent
on the heating speed. This can happen if the nucleation time
between two mesoglobular states with different average cluster
sizes is much longer than the typical measurement time involved.
Thus, even though one of such mesoglobular states would have the lowest
free energy at given thermodynamic parameters, the system can in principle
be trapped for a rather long time in another such state which happens
to be merely metastable.

It is worthwhile to emphasise that
in solutions of a biopolymer, such as e.g. a protein, all chains would have
{\it exactly} the same structure because they are produced 
by the unique rules from the same genetic code.
However, synthesis normally results in a 
mixture of chains with somewhat varying lengths and sequences 
and the presence of various defects. 
Thus, it would be interesting to
investigate the influence of weak imperfections remaining
after applying physical methods such as fractionation and centrifugation
on the monodispersity of the mesoglobules in solution.
Technically, this requires
to perform a quenched disorder averaging: first over the
identical random sequences, and then also to permit randomness
in the structure of each chain in solution.
Replica techniques \cite{MezPar} are commonly
adopted for such purpouses and we believe they may lead to further
progress in studying the current problem and its possible variations.

Finally, we hope that 
mesoglobular structures may find a number of interesting industrial
applications as their size distribution may be well controlled.
Another potential application of these results would be in learning
how to facilitate folding and suppress aggregation of
proteins {\it in vitro}.

\acknowledgments

The authors acknowledge interesting discussions with
Professor M.~Gitterman, Professor A.Yu.~Grosberg
and our colleagues Dr A.V.~Gorelov and Professor K.A.~Dawson.
This work was supported by grant SC/99/186 from Enterprise Ireland.



\begin{figure}
\caption{ \label{fig:mh4}
Plot of the mean squared cluster size, $R^2_{cl}$,
vs the second virial coefficient,
$u^{(2)}$, for different cluster states.
This data is obtained for open homopolymers with $N = 18$, $M = 4$ and
$L = 10$.
The dashed vertical line I corresponds to the transition point at which
the free energies of states $4\times 1$ (single chains) and
$1\times 4$ (aggregate) become equal.
}
\end{figure}


\begin{figure}
\caption{ \label{fig:mcp_1a1b_phd}
Phase diagram for solution of $M = 4$ $(ab)_6$ heteropolymers in terms of the
amphiphilicity, $\Delta$, and the mean second virial coefficient,
$\bar{u}^{(2)}$. The linear box size is $L = 8$. Here and below
the transition curves have been determined by the condition of free energy
equality on corresponding minima as in previous figure.
}
\end{figure}


\begin{figure}
\caption{ \label{fig:mcp_3a3b_phd}
Phase diagram for solution of $M = 4$ $(a_3b_3)_2$ heteropolymers
in terms of the amphiphilicity, $\Delta$, and the mean second
virial coefficient, $\bar{u}^{(2)}$. Here the linear box size is $L = 8$.
}
\end{figure}


\begin{figure}
\caption{ \label{fig:nclvt}
Time dependence of the total number of clusters $n_{cl}$ on the
approach to final equilibrium for different polymer sequences. 
1 Monte Carlo sweep is defined as $NM$ attempted Monte Carlo moves.
Here $L=60$, $N=24$, $M=20$, $\chi_{aa}=1$, and $\chi_{bb}=-0.2$.
}
\end{figure}


\begin{figure}
\caption{ \label{fig:SnaPgoodNbad}
Snapshots of typical polymer conformations from Monte Carlo simulation:
Fig. a --- for the good solvent condition for sequence $(a_3 b_3)_4$, and
Fig. b --- the single aggregate for diblock sequence $a_{12} b_{12}$.
All parameters here are as in Fig.~\ref{fig:nclvt}. Black and white
circles correspond to hydrophobic and hydrophilic monomer units respectively.
}
\end{figure}

\begin{figure}
\caption{ \label{fig:SnaPmeso}
Snapshots of typical polymer conformations from Monte Carlo simulation
for mesoglobules of different heteropolymers. Figs.~a-d correspond to
sequences: $(a_2 b_2)_6$, $(a_3 b_3)_4$,  
$b_3 a b_2 a_3 b a_4 b a_3 b_2 a b_3$ and $a_{12} b_{12}$ respectively
(these are also called as sequences s1, s2, s6 and s4 in Figs.~\ref{fig:MclHys}
and \ref{fig:RclHys}).
All parameters here are as in Fig.~\ref{fig:nclvt}. Black and white
circles correspond to hydrophobic and hydrophilic monomer units respectively.
}
\end{figure}


\begin{figure}
\caption{ \label{fig:MclHys}
Histogram of the number of chains (mass) $M_{cl}$ constituting mesoglobules
for different heteropolymer sequences. These results have been obtained
by analyzing data for the ensemble size (the number of initial conditions) 
$Q=1000$ after equilibration time $4\cdot 10^6$ MC sweeps
(the last time point in Fig.~\ref{fig:nclvt}).
All parameters here are as in Fig.~\ref{fig:nclvt}. Note that for some
sequences there are large populations of single globules ($M_{cl}=1$) and 
single aggregate ($M_{cl}=20$).
}
\end{figure}

\begin{figure}
\caption{ \label{fig:RclHys}
Histogram of the squared radius of gyration $R_{cl}^2$ of mesoglobules
for different heteropolymer sequences. 
All parameters here are as in Fig.~\ref{fig:MclHys}.
}
\end{figure}

%

\begin{table}
\caption{ \label{tab:many_hom}
Values of the specific free energy, $a = {\cal A}/MN$, at various
minima for the system of $M = 12$ open homopolymers in a box with $L = 20$
for various values of the second virial coefficient, $u^{(2)}$.
The value of the global (deepest) minimum of the free energy in
each line is printed in {\bf bold} face.
The horizontal line here corresponds to the transition point, i.e.
equality condition of free energies for states $12\times 1$ (separate
chains) and $1\times 12$ (aggregate).
}
\vskip 3mm
{\small
\begin{tabular}{|r||llllll|ll|}
$u^{(2)}$ & $12\times 1$ & $6\times 2$ & $4\times 3$ & $3\times 4$ &
            $2\times 6$ & $1\times 12$ & $1,11$ & $1,1,10$ \\
\hline \hline
$-12$     &${\bf -1.028}$& $-0.843$  & $-0.815$   & $-0.814$   &
            $-0.832$ & $-0.886$ & $-0.890$ & $-0.896$ \\
$-13$     &${\bf -1.378}$& $-1.206$  & $-1.185$   & $-1.188$   &
            $-1.210$ & $-1.270$ & $-1.271$ & $-1.274$ \\
$-14$     &${\bf -1.759}$& $-1.604$  & $-1.589$   & $-1.596$   &
            $-1.623$ & $-1.688$ & $-1.685$ & $-1.685$ \\
$-15$     &${\bf -2.172}$ & $-2.035$   & $-2.027$    & $-2.038$    &
            $-2.069$ & $-2.140$ & $-2.132$ & $-2.127$  \\
\hline
$-16$     & $-2.615$     & $-2.499$  & $-2.498$   & $-2.513$   &
            $-2.548$ &${\bf -2.626}$& $-2.615$ & $-2.606$ \\
$-18$     & $-3.595$     & $-3.523$  & $-3.539$   & $-3.562$   &
            $-3.607$ &${\bf -3.695}$& $-3.676$ & $-3.659$ \\
$-20$     & $-4.696$      & $-4.676$   & $-4.709$    & $-4.742$    &
            $-4.797$ &${\bf -4.897}$& $-4.867$ & $-4.840$ \\
$-25$     & $-7.982$      & $-8.113$   & $-8.198$    & $-8.258$    &
            $-8.341$ &${\bf -8.472}$& $-8.416$ & $-8.386$ \\
\end{tabular}}
\end{table}


\begin{table}
\caption{ \label{tab:mcp_1a1b}
Values of the specific free energy, $a = {\cal A}/MN$, at various
minima for the system of $M = 12$ $(ab)_6$ heteropolymers for various values
of the mean second virial coefficient, $\bar{u}^{(2)}$.
Here and below the linear system size and the
amphiphilicity are equal to $L = 20$ and $\Delta = 30$.
The horizontal lines here and below delimit the broad region of stable
mesoglobules, i.e. where the global free energy minimum is reached on
states other than $12\times 1$ or $1\times 12$.
}
\vskip 3mm
{\footnotesize
\begin{tabular}{|r||llllll|lll|}
$\bar{u}^{(2)}$ & $12\times 1$ & $6\times 2$ & $4\times 3$ & $3\times 4$ &
            $2\times 6$ & $1\times 12$ & $1,11$ & $2,10$ & $1,1,10$ \\
\hline \hline
$-5$      & ${\bf -4.25}$ & $-3.70$  & $-3.25$    & $2.89$     &
            $-2.33$  & ---  & $-4.17$  & $-3.46$  & $-3.56$  \\
$-10$     & ${\bf -7.08}$ & $-6.78$  & $-6.42$    & $-6.09$    &
            $-5.57$  & $-4.54$ & $-4.88$  & $-5.15$  & $-5.20$  \\
\hline
$-15$     & $-10.31$ & ${\bf -10.31}$  & $-10.05$    & $- 9.79$    &
            $- 9.32$ & $- 8.34$ & $- 8.63$ & $- 8.90$ & $- 8.90$ \\
$-20$     & $-13.96$ & ${\bf -14.29}$  & $-14.16$    & $-13.97$    &
            $-13.58$ & $-12.69$ & $-12.91$ & $-13.17$ & $-13.12$ \\
$-25$     & $-18.04$ & $-18.74$  & ${\bf -18.75}$    & $-18.64$    &
            $-18.35$ & $-17.58$ & $-17.72$ & $-17.97$ & $-17.85$ \\
$-30$     & $-22.61$ & $-23.69$  & ${\bf -23.86}$    & $-23.83$    &
            $-23.65$ & $-23.02$ & $-23.07$ & $-23.29$ & $-23.12$ \\
$-35$     & $-27.73$ & $-29.18$  & $-29.51$    & ${\bf -29.57}$    &
            $-29.50$ & $-29.03$ & $-28.99$ & $-29.19$ & $-28.94$ \\
$-40$     & $-33.87$ & $-35.27$  & $-35.75$    & $-35.91$    &
            ${\bf -35.95}$ & $-35.64$ & $-35.51$ & $-35.68$ & $-35.37$ \\
$-45$     & $-40.41$ & $-42.47$  & $-42.65$    & $-42.90$    &
            ${\bf -43.04}$ & $-42.88$ & $-42.67$ & $-42.81$ & $-42.45$ \\
$-50$     & $-47.64$ & $-50.05$  & $-50.65$    & $-50.68$    &
            ${\bf -50.83}$ & $-50.81$ & $-50.52$ & $-50.63$ & $-50.21$ \\
\hline
$-55$     & $-55.57$ & $-58.36$  & $-58.69$    & $-59.05$    &
            $-59.34$ & ${\bf -59.46}$ & $-59.08$ & $-59.17$ & $-58.69$ \\
\end{tabular}}
\end{table}


\begin{table}
\caption{ \label{tab:mcp_2a2b}
Values of the specific free energy, $a = {\cal A}/MN$, at various
minima for the system of $M = 12$ $(a_2 b_2)_3$ heteropolymers for
various values of the mean second virial coefficient, $\bar{u}^{(2)}$.}
\vskip 3mm
{\footnotesize
\begin{tabular}{|r||llllll|lll|}
$\bar{u}^{(2)}$ & $12\times 1$ & $6\times 2$ & $4\times 3$ & $3\times 4$ &
            $2\times 6$ & $1\times 12$ & $1,11$ & $2,10$ & $1,1,10$ \\
\hline \hline
$-5$      & ${\bf -4.98}$ & $-4.77$  & $-4.53$    & $4.31$     &
            $-3.94$  & $-3.20$ & $-3.44$  & $-3.63$  & $-3.67$  \\
\hline
$-10$     & $- 7.75$ & ${\bf - 7.80}$  & $- 7.65$    & $- 7.47$    &
            $- 7.16$ & $- 6.47$ & $- 6.66$ & $- 6.85$ & $- 6.85$ \\
$-15$     & $-10.91$ & ${\bf -11.25}$  & $-11.21$    & $-11.09$    &
            $-10.85$ & $-10.23$ & $-10.37$ & $-10.55$ & $-10.50$ \\
$-20$     & $-14.47$ & $-15.14$  & ${\bf -15.21}$    & $-15.17$    &
            $-15.00$ & $-14.48$ & $-14.55$ & $-14.73$ & $-14.62$ \\
$-25$     & $-18.45$ & $-19.48$  & $-19.69$    & ${\bf -19.72}$    &
            $-19.64$ & $-19.24$ & $-19.24$ & $-19.39$ & $-19.22$ \\
$-30$     & $-22.91$ & $-24.30$  & $-24.66$    & $-24.77$    &
            ${\bf -24.79}$ & $-24.42$ & $-24.43$ & $-24.57$ & $-24.34$ \\
$-35$     & $-27.93$ & $-29.66$  & $-30.16$    & $-30.36$    &
            ${\bf -30.48}$ & $-30.35$ & $-30.18$ & $-30.29$ & $-30.00$ \\
\hline
$-40$     & $-34.02$ & $-36.03$  & $-36.27$    & $-36.55$    &
            $-36.76$ & ${\bf -36.77}$ & $-36.52$ & $-36.51$ & $-36.27$ \\
$-45$     & $-40.51$ & $-42.82$  & $-43.27$    & $-43.42$    &
            $-43.57$ & ${\bf -43.85}$ & $-43.53$ & $-43.49$ & $-43.20$ \\
\end{tabular}}
\end{table}


\begin{table}
\caption{ \label{tab:mcp_3a3b}
Values of the specific free energy, $a = {\cal A}/MN$, at various
minima for the system of $M = 12$ $(a_3 b_3)_2$ heteropolymers for
various values of the mean second virial coefficient, $\bar{u}^{(2)}$.}
\vskip 3mm
{\footnotesize
\begin{tabular}{|r||llllll|lll|}
$\bar{u}^{(2)}$ & $12\times 1$ & $6\times 2$ & $4\times 3$ & $3\times 4$ &
            $2\times 6$ & $1\times 12$ & $1,11$ & $2,10$ & $1,1,10$ \\
\hline \hline
$0$       & ${\bf -3.02}$ & $-2.81$  & $-2.61$    & $2.45$     &
            $-2.18$  & $-1.65$ & $-1.83$  & $-1.97$  & $-2.01$  \\
\hline
$-5$      & $-5.38$ & ${\bf -5.39}$  & $-5.27$    & $5.14$     &
            $-4.91$  & $-4.40$ & $-4.55$  & $-4.69$  & $-4.69$  \\
$-10$     & $- 8.12$ & ${\bf - 8.37}$  & $- 8.34$    & $- 8.26$    &
            $- 8.08$ & $- 7.63$ & $- 7.73$ & $- 7.89$ & $- 7.83$ \\
$-15$     & $-11.23$ & $-11.78$  & ${\bf -11.85}$    & $-11.83$    &
            $-11.71$ & $-11.33$ & $-11.38$ & $-11.51$ & $-11.42$ \\
$-20$     & $-14.74$ & $-15.60$  & $-15.80$    & ${\bf -15.840}$    &
            $-15.79$ & $-15.51$ & $-15.49$ & $-15.60$ & $-15.46$ \\
$-25$     & $-18.66$ & $-19.87$  & $-20.19$    & $-20.31$    &
            ${\bf -20.34}$ & $-20.16$ & $-20.07$ & $-20.17$ & $-19.97$ \\
$-30$     & $-23.05$ & $-24.61$  & $-25.070$    & $-25.26$    &
            ${\bf -25.39}$ & $-25.32$ & $-25.16$ & $-25.24$ & $-24.98$ \\
\hline
$-35$     & $-28.00$ & $-29.89$  & $-30.48$    & $-30.75$    &
            $-30.97$ & ${\bf -31.03}$ & $-30.78$ & $-30.85$ & $-30.54$ \\
\end{tabular}}
\end{table}


\begin{table}
\caption{ \label{tab:mcp_rand}
Values of the specific free energy, $a = {\cal A}/MN$, at various
minima for the system of $M = 12$ heteropolymers with the sequence
$b_3 a_2 b a_2 b a b a$ for various values
of the mean second virial coefficient, $\bar{u}^{(2)}$.}
\vskip 3mm
{\footnotesize
\begin{tabular}{|r||llllll|lll|}
$\bar{u}^{(2)}$ & $12\times 1$ & $6\times 2$ & $4\times 3$ & $3\times 4$ &
            $2\times 6$ & $1\times 12$ & $1,11$ & $2,10$ & $1,1,10$ \\
\hline \hline
$-10$     & ${\bf - 7.78}$ & $- 7.77$  & $- 7.61$    & $- 7.43$    &
            $- 7.10$ & $- 6.40$ & $- 6.60$ & $- 6.80$ & $- 6.80$ \\
\hline
$-15$     & $-10.93$ & ${\bf -11.23}$  & $-11.16$    & $-11.03$    &
            $-10.77$ & $-10.14$ & $-10.29$ & $-10.43$ & $-10.48$ \\
$-20$     & $-14.49$ & $-15.11$  & ${\bf -15.17}$    & $-15.11$    &
            $-14.92$ & $-14.37$ & $-14.46$ & $-14.53$ & $-14.64$ \\
$-25$     & $-18.47$ & $-19.44$  & $-19.63$    & ${\bf -19.65}$    &
            $-19.54$ & $-19.11$ & $-19.12$ & $-19.12$ & $-19.29$ \\
$-30$     & $-22.92$ & $-24.26$  & $-24.59$    & ${\bf -24.69}$    &
            $-24.68$ & $-24.37$ & $-24.30$ & $-24.22$ & $-24.45$ \\
$-35$     & $-27.91$ & $-29.60$  & $-30.08$    & $-30.27$    &
            ${\bf -30.36}$ & $-30.18$ & $-30.03$ & $-29.97$ & $-30.15$ \\
$-40$     & $-33.97$ & $-35.96$  & $-36.19$    & $-36.45$    &
            ${\bf -36.63}$ & $-36.59$ & $-36.36$ & $-36.13$ & $-36.46$ \\
\hline
$-45$     & $-40.47$ & $-42.75$  & $-43.26$    & $-43.31$    &
            $-43.59$ & ${\bf -43.67}$ & $-43.37$ & $-43.06$ & $-43.44$ \\
$-50$     & $-47.67$ & $-50.27$  & $-50.99$    & $-51.18$    &
            $-51.26$ & ${\bf -51.46}$ & $-51.09$ & $-50.71$ & $-51.15$ \\
\end{tabular}}
\end{table}


\end{document}